\documentclass[11pt]{article}
\usepackage{amsmath}
\usepackage{amssymb}
\usepackage{graphics}
\usepackage{epsf,graphicx,rotating,color}
\usepackage{bm,bbm}
\usepackage{amsfonts}

\begin{document}
\title{\bf About Radiation Reaction with Force Approach}
\author{G.V. L\'opez\footnote{gulopez@udgserv.cencar.udg.mx}\\
\small \empty{Departamento de F\'{i}sica, Universidad de Guadalajara,}\\
\small \emph{Blvd. Marcelino Garc\'{i}a Barragan y Calzada Ol\'{i}mpica,}
 \small \emph{C.P. 44200 Guadalajara, Jalisco, Mexico}}
\date{\today}
\maketitle
\begin{center}
{\large\bf Abstract}
\end{center}
\vskip2pc
The difficulty of usual approach to radiation reaction is pointed out , and a possible approach,  based on the force acting to the charged particle
which produces the acceleration itself, is presented. This approach brings about an expression such that acceleration is zero whenever the external force is zero. \\ \\
\centerline{{\bf PACS:} 41.60.-m, 41.60.Ap, 41.20.-q, 41.20.Jb}
\newpage
\section{ Introduction}
Radiation of electromagnetic waves due to the acceleration of charged particles is a very well known classical phenomenon predicted and explained by Maxwell's equations [1,2], and this phenomenon has been used for practical proposes World wide [3,4,5,6]. This radiation, of course, implies dissipation of energy and damping motion of the charged particle, and the known modification to the equation of motion to take into account this damping effect are the so called Abraham-Lorentz (non relativistic case [7,8]) and Lorentz-Dirac (relativistic case [9]) equations. These equations have the particularity that even if the external force (responsible of the acceleration of the charged particle) is zero, an acceleration of the particle still exists. On the other hand, one main experimental fact needed to take into account  is that this  this radiation of electromagnetic waves due to acceleration of charges  disappears as soon as the acceleration disappears, and this acceleration disappears as soon the external force is zero. This implies that radiation force (damping force associated to emission of electromagnetic waves) must be a function of this external force. From this point of view, Abraham-Lorentz-Dirac formulation of radiation damping is not totally satisfactory [10,11,12] since one still have solutions of their equations with acceleration of the charge particle, even if the total external force is zero. In this paper, one considers a different point of view of this radiation force  and arrives to an expression that takes into account the experimental fact, and it might worth to study it. In this paper,  one considers  linear radiation firstly, and circular radiation secondly.
\section{ Linear radiation force}

As it is well know, the total power radiated in a linear acceleration motion  of a charged particle with charge "e" as a function of the external force; ${\bf F}$ with magnitude $F$, is (CGS units) 
\begin{equation}
P=\frac{2}{3}\frac{e^2F^2}{m^2c^3}\ ,
\end{equation}       
where "c" is the speed of light ($c\approx 3\times 10^8m/s$ ) and $m$ is the mass of the charge. This means that the energy lost by the charged particle from the time $t=0$ (time at which the external force in on) to the time $t$ is
\begin{equation}\label{eloss}
U=\frac{2e^2}{3m^2c^3}\int_0^tF^2dt\ .
\end{equation}
Assume that this energy lost is due to a non conservative radiation force, ${\bf F}_{rad}$, and that the charged particle travels from the point ${\bf x}_0$,at the time $t=0$,  to the point ${\bf x}$, at the time $t$. Then one gets that
\begin{subequations}
\begin{equation}
U=\lambda_0\int_0^tF^2dt=\int_{{\bf x}_0}^{\bf x}{\bf F}_{rad}\cdot d{\bf x}\ ,
\end{equation}
where one has made the definition
\begin{equation}
\lambda_0=\frac{2e^2}{3m^2c^3}.
\end{equation}
\end{subequations}
Since one has that $d{\bf x}={\bf v} dt$, where ${\bf v}$ represents the velocity of the charged particle, one obtains 
\begin{equation}
\int_0^t\big(\lambda_0F^2-{\bf F}_{rad}\cdot{\bf v}\bigr)dt=0\ ,
\end{equation}
for any time interval $[0,t]\subset\Re$. This means that the integrand must be zero, and being $\theta_r$ the angle between the the velocity of the charged particle and the radiation force (${\bf F}_{rad}\cdot{\bf v}=F_{rad}v\cos\theta_r$), one has the magnitude of the radiation force given by
\begin{equation}
F_{rad}=\frac{\lambda_0F^2}{v\cos\theta_r}.
\end{equation}
This force must be responsible of the damping motion of charge motion. Therefore, its direction must be opposite to the direction of the velocity of the charge. Thus, one must have that $\theta_r=\pi$, and ,if $\hat {\bf n}=\hat{\bf v}/v$ is the unitary vector in the direction of the velocity of the charge, one also must have that
\begin{equation}
{\bf F}_{rad}=-\frac{\lambda_0F^2}{v} \hat{\bf n}\ ,
\end{equation} 
or
\begin{equation}
{\bf F}_{rad}=-\frac{\lambda_0F^2}{v^2}{\bf v}.
\end{equation}
Then, the modified equation of motion of the charged particle under an external force ${\bf F}$ and damping radiation force ${\bf F}_{rad}$ is
\begin{equation}\label{meq}
\frac{d(m\gamma{\bf v})}{dt}={\bf F}-\frac{\lambda_0 F^2}{v^2}{\bf v}.
\end{equation}
For a charged particle moving in the z-direction, ${\bf v}=(0,0,v)$, and having an external force also in the same direction, ${\bf F}=(0,0,F)$, equation (\ref{meq}) is reduced to  the following Newton like equation of motion
\begin{equation}\label{d1meg}
\frac{d (mv)}{dt}=F\left(1-\frac{\lambda_0F}{v}\right)\left(1-\frac{v^2}{c^2}\right)^{3/2}.
\end{equation}
For example, if the force is constant, equation (\ref{d1meg}) can be solved, and its solution can be written as
\begin{equation}\label{meq}
f(\beta,F)-f(\beta_0,F)+g(\beta,F)-g(\beta_0,F)=\frac{Ft}{mc},
\end{equation}
where $\beta$ and $\beta_0$ have been defined as $\beta=v/c$ and $\beta_0=v_0/c$, with $v_0$ indicating the initial condition of motion.  The functions $f$ and $g$ have been defined as
\begin{subequations}
\begin{equation}
f(\beta,F)=
\frac{\displaystyle\left[1-\left(\frac{\lambda_0F}{c}\right)^2\right]\left(\beta-\frac{\lambda_0F}{c}\right)+\frac{2\lambda_0F}{c}}
{\displaystyle\left[1-\left(\frac{\lambda_0F}{c}\right)^2\right]\sqrt{1-\beta^2}}
\end{equation}
and
\begin{equation}
g(\beta,F)=-\frac{\lambda_0F/c}{\displaystyle\left[1-\left(\frac{\lambda_0F}{c}\right)\right]^{3/2}}
\ln\left[\frac{2\left(1-\left(\frac{\lambda_0F}{c}\right)^2\right)-\frac{2\lambda_0F}{c}\left(\beta-\frac{\lambda_0F}{c}\right)+2\sqrt{1-\left(\frac{\lambda_0F}{c}\right)^2}~\sqrt{1-\beta^2}}{\displaystyle \beta-\frac{\lambda_0F}{c}}\right]\ .
\end{equation}
\end{subequations}
For the non relativistic limit, $\beta\ll 1$, one gets from (\ref{meq}) the following expression
\begin{equation}\label{nrmeq}
\beta-\beta_0+\frac{\lambda_0F}{c}\ln\left[\frac{\beta-\lambda_0F/c}{\beta_0-\lambda_0F/c}\right]=\frac{Ft}{mc}\ .
\end{equation}
From the expressions (\ref{meq}) or (\ref{nrmeq}) one can sees that for $F=0$, one gets $\beta=\beta_0$, that is, there is not acceleration of any kind. Therefore, there is not radiation, and the charged particle will travel at constant velocity. For the non relativistic case, Figure 1 shows the difference between non radiation motion, $\beta_o=\beta_0+Ft/mc$, and the expression (\ref{nrmeq}). This difference increases rapidly and reaches an asymptotic behavior (the particle loses energy with time (\ref{eloss}), but the external force feeds energy with time).  
\begin{figure}[H]
\includegraphics[scale=0.60]{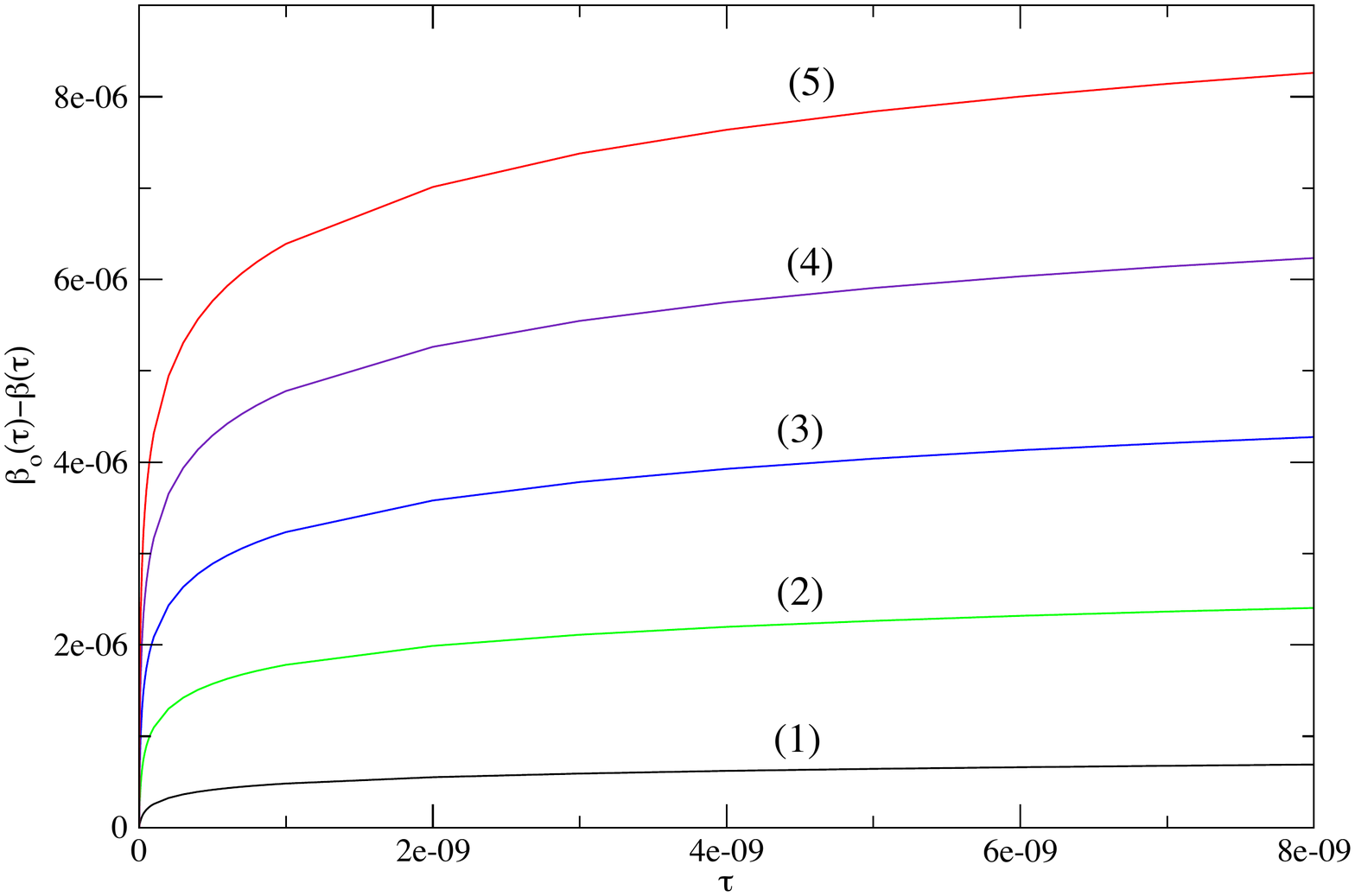}
\centering
    \caption{ Radiation effect: (1)  }
\end{figure}
\section{ Circular radiation force}
In this case, the total power emitted by the charged particle in terms of the magnitude of the external force is
\begin{equation}
P=\frac{2e^2}{3m^2c^3}\frac{F^2}{\gamma^2},
\end{equation}
where $\gamma=(1-\beta^2)^{-1/2}$, and the total energy emitted in the interval of time $[o,t]\subset\Re$ is
\begin{equation}
U=\lambda_0\int_0^t\frac{F^2}{\gamma^2}dt=\int_{{\bf x}_0}^{\bf x}{\bf F}_{rad}\cdot d{\bf x},
\end{equation}
where one has made the assumption that this energy is due to the work done by a radiation force to move the charge particle from the initial point ${\bf x}_0$ at the time $t=0$, to the point ${\bf x}$ at the time $t$. Therefore, using the relation $d{\bf x}={\bf v}dt$ as before, one gets (using the same arguments) that the magnitude of the radiation force is 
\begin{equation}
F_{rad}=\frac{\lambda_0F^2}{v\gamma\cos\theta_r},
\end{equation} 
where $\theta_r$ is the angle between the velocity of the charged particle, ${\bf v}$, and the radiation force, ${\bf F}_{rad}$. Once again, since the direction of the radiation force must be opposite to the velocity of the charged particle, one must have that $\theta_r=\pi$, and the radiation force must be in the direction $\hat{\bf n}={\bf v}/v$. Thus, the radiation force can be written as
\begin{equation}
{\bf F}_{rad}=-\frac{\gamma_0F^2}{v^2\gamma^2}{\bf v}.
\end{equation}  
In this way, the relativistic equation of motion of the charged particle with rest mass $m$ and external force ${\bf F}$ is
\begin{equation}\label{rcirc}
\frac{d(\gamma m{\bf v})}{dt}={\bf F}-\frac{\lambda_0F^2}{v^2\gamma^2}{\bf v}.
\end{equation}
Now, since for circular motion one has that ${\bf v}\cdot d{\bf v}/dt=0$, this equation can be written as the following Newton like equation of motion
\begin{equation}\label{mcirc}
\frac{d(m{\bf v})}{dt}=\biggl({\bf F}-\frac{\lambda_0F^2}{v^2\gamma^2}{\bf v}\biggr)\frac{1}{\gamma}\ .
\end{equation}
For example, assume one has a constant magnetic in the z-direction and the charge particle is moving on the x-y plane, ${\bf v}=(v_x,v_y,0)$. Then the external force is
\begin{equation}
{\bf F}=\frac{eB_0}{c}(v_y,-v_x,0),
\end{equation} 
and (\ref{mcirc}) can be written as the following dynamical system
\begin{subequations}
\begin{eqnarray}\label{dycirc}
& & \dot x=v_x\quad\quad \dot v_x=\frac{1}{\gamma }\biggl(\sigma v_y-\frac{\lambda_0F^2}{mv^2\gamma^2}v_x\biggr)\\ \nonumberÊ\\
& & \dot y=v_y\quad\quad \dot v_y=\frac{1}{\gamma }\biggr(-\sigma v_x-\frac{\lambda_0F^2}{mv^2\gamma^2}v_y\biggr)\ ,
\end{eqnarray}
\end{subequations}
where one has $\sigma=eB_0/mc$. For $\lambda_0=0$  and non relativistic particle motion ($\gamma\approx 1$), one knows that the charge particle will describe a circular motion around the magnetic field with a frequency $\sigma$ and a radius $R=v_0/\sigma$ (being $v_0$ the initial velocity of the charged particle on the plane).  Figure 2 shows the difference on the trajectories between non radiation force and with radiation force $S_x=\sqrt{(\delta x)^2+(\delta y)^2}$. As stronger the external force is, bigger is the difference, and the oscillations come from higher and lower acceleration of the charged particle. These maxima and minima are different for different external force due to radiation dissipation. 
\begin{figure}[H]
\includegraphics[scale=0.60]{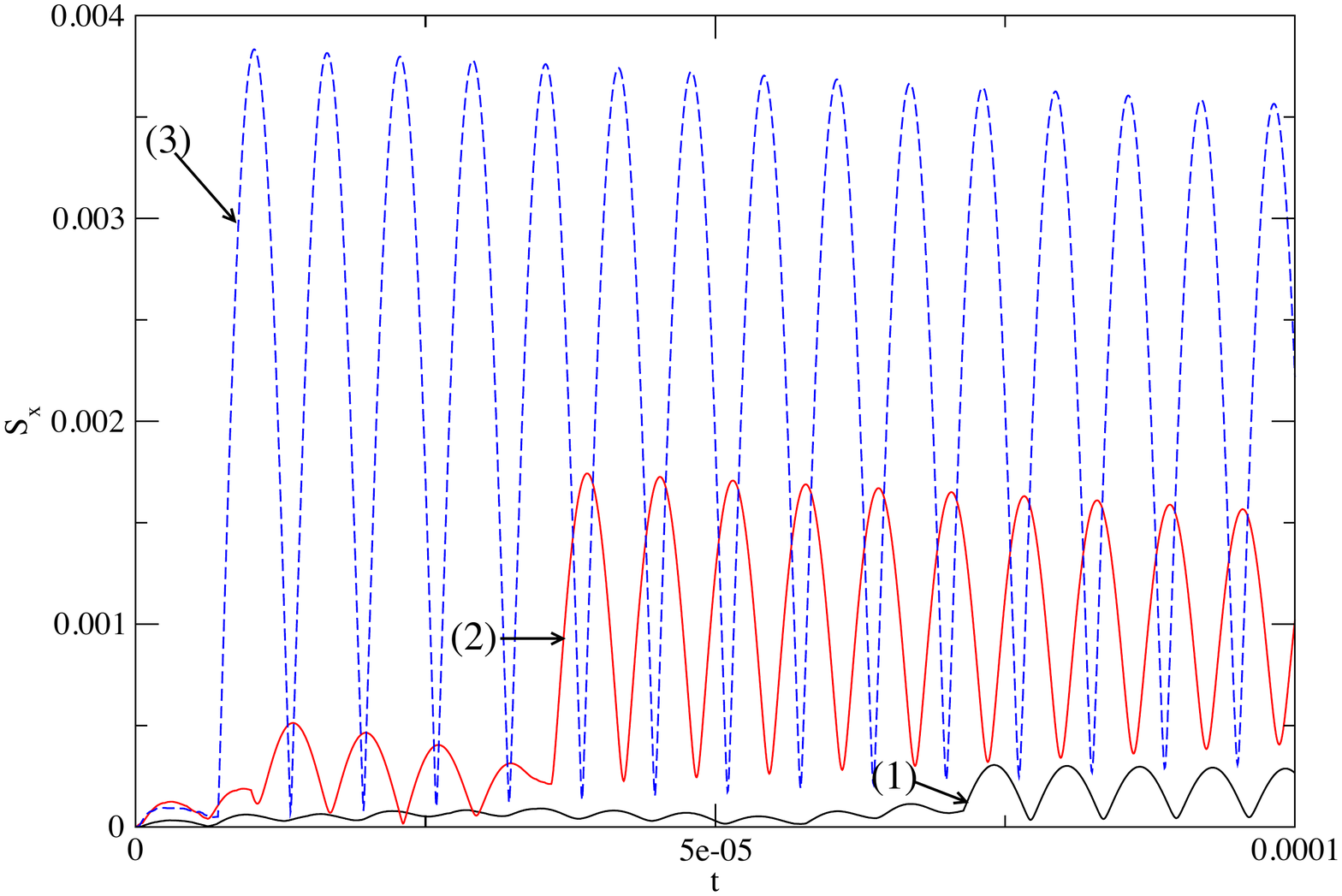}
\centering
    \caption{ Radiation effect: (1) F=1Dina, (2) F=5 Dinas, (3) F=8 Dinas. $\sigma=10^6Hz$, $\lambda_0=10^8$.  }
\end{figure}

\section{Conclusions}
 Under the condition that the radiation force must be a function of the external force and to be zero whenever the external force be zero, an expression
 for the radiation force was given for the linear and circular motion of a charged particle. %


\begin{thebibliography}{99}
%
   \bibitem{Jak}  J.D. Jackson {\it Classical Electrodynamics}, John Wiley\& Sons Inc., ( 1962).
   \bibitem{Landau} L.D. Landau and E.M. Lifshitz, {\it The Classical Theory of Fields}, Pergamon Press, (1971).
   \bibitem{Saldin} E.L. Saldin, E.A. Scheidaniller, and M.V. Yurkov, {\it The Physics of Free electron Laser}, Springer, Berlin, (1999).
  \bibitem{Elder} F.R. Elder, R.V. Langmuir, and H.C. Pollock, {Phys. Rev.}, {\bf 74}, (1948), 52.
   \bibitem{Strong} A. Strong, I.V. Moskalenko, and O. Reimer, {ApJ}, (2004), 613.
   \bibitem{Piazza} A. Di Piazza, K.Z. Hatsagortsyan, and C.H. Keitel, {Phys. Rev. Lett.}, {\bf 102}, (2009), 254802-1.
   \bibitem{Abraham}M. Abraham, {\it Theorie de Electrizit\"at}, Teubner, Leipzig, Vol. II  (1905).
   \bibitem{Lorentz} H.A. Lorentz, {\it The Theory of Electrons}, Teubner, Leipzig, (1909).
   \bibitem{Dirac} P.A.M. Dirac, Proc. Roy. Soc. London A{\bf 167} (2003), 148.
   \bibitem{Comay} E. Comay, {Found. Phys.}, {\bf 23}, (1993),1121.
   \bibitem{Valentini} A. Valentini, {Phys. Rev. Letts.},  {\bf 61},  (1988), 1903.
   \bibitem{Spohn} H. Spohn,{\it Dynamics of Charge Particles and Their Radiation Fields},  Cambridge University Press, (2004).   
%
  \end{thebibliography}
\end{document}